%% file: paper.tex
\documentclass[10pt,a4paper]{extarticle}
\usepackage[top=2cm, bottom=2cm, left=2.5cm, right=2.5cm, a4paper]{geometry}

\RequirePackage{fix-cm}
\RequirePackage{hyperref}

\hypersetup{
    colorlinks=false,
    pdfborder={0 0 0},
}

\usepackage{lettrine}

\usepackage{amsmath} 
\usepackage{amssymb}

\DeclareMathOperator{\Tr}{Tr} 

\usepackage{subfigure}
\usepackage{subfigmat}

\usepackage{pgf}  
\usepackage{graphicx}
\graphicspath{ {./pics/} }

\usepackage{pgfplots}
\pgfplotsset{compat=newest}
\pgfplotsset{plot coordinates/math parser=false}

\usepackage{tikz}
\usetikzlibrary{shapes,arrows}
\tikzset{
  treenode/.style = {align=center, inner sep=0pt, text centered,
    font=\sffamily},
  arn_x/.style = {treenode, rectangle, draw=black,
    minimum width=2cm, minimum height=.5cm,inner sep=.1cm}%
}

\usepackage{cleveref} 

\usepackage{fancyhdr}
\usepackage[figwidth=5cm]{todonotes}
\usepackage{soul}

\usepackage[acronym, nonumberlist, style=super]{glossaries} 
\glsdisablehyper
\input{glossary.tex}
\makeglossaries

\usepackage{setspace}       
\usepackage{lineno}         

\newlength{\singlepic}
\newlength{\doublepic}

\title{Trajectory Clustering, Modelling, and Selection with the focus on Airspace Protection}

\author{
Willem J. Eerland \thanks{Postgraduate researcher, Transportation Research Group, \texttt{w.j.eerland@soton.ac.uk}} \ and
Simon Box \thanks{Lecturer, Transportation Research Group} \\ 
{\normalsize\itshape
 University of Southampton, Southampton, England SO17 1BJ, United Kingdom}\\
}

\begin{document}


\maketitle

\begin{abstract}
Take-off and landing are the periods of a flight where aircraft are most vulnerable to a ground based rocket attack by terrorists. While aircraft approach and depart from airports on pre-defined flight paths, there is a degree of uncertainty in the trajectory of each individual aircraft. Capturing and characterizing these deviations is important for accurate strategic planning for the defence of airports against terrorist attack.
A methodology is demonstrated whereby approach and departure trajectories to a given airport are characterized statistically from historical data. It uses a two-step process of first clustering to extract the common trend, and then modelling uncertainty using \acrfull{gps}. Furthermore it is shown that this approach can be used to either select probabilistic regions of airspace where trajectories are likely and - if required - can automatically generate a set of representative trajectories, or select key trajectories that are both likely and critically vulnerable.
An evaluation of the methodology is demonstrated on an example data-set collected by the ground radar at an airport. The evaluation indicates that $99.8\%$ of the calculated footprint underestimates less than $5\%$ when replacing the original trajectory data with a set of representative trajectories.
\end{abstract}


\section{Introduction}

\lettrine[nindent=0pt]{E}{xtracting} patterns from data is an active field for both research and industry, ranging from tracking traffic to making predictions on the financial market. It allows for objects to be clustered when following a similar trend. And by identifying the generic response, an attempt can be made to explain these reactions. When it comes to air traffic, the reason is well known. As an aircraft enters controlled airspace, such as near an airport, they follow the instructions of air traffic control, who guide them along pre-designated paths.

In the real-world, these pre-designated paths, also called flight paths, have more resemblance to corridors. For this reason, the current methodology on calculating noise contours around civil airports in Europe, uses several sub-tracks to model the dispersion along a single flight path \cite{ECAC2005}. When using the \gls{inm} \cite{boeker2008integrated} to calculate the noise contours, either the sub-tracks or dispersion will have to be supplied by the user.
Nowadays, with the air traffic increasing in volume, the introduction of new technologies and procedures being developed under the names \emph{NextGen} and \emph{SESAR}, the dispersion along the flight paths is more susceptible to change than ever before. A step-by-step guide to obtain a number of representative trajectories from historical data (over any given time-frame or conditions), is able to reduce the computational load in any subsequent analysis, without losing integrity and with limited effort for the user.

While noise can literally keep people awake at night, it is nowhere near as vital as securing the infrastructure. More specifically, protecting aircraft against the threat from \gls{rpgs}, that can hit a moving target up to a distance of 300 metres \cite{aviationtoday}. The threat is real, as actions in the past, such as the heightened security around Heathrow back in 2003 have shown \cite{guardian-missile-terror}.
This is one of the reasons governments, airports and airliners alike perform much strategic planning to defend aircraft from terrorists attacks.
%
In the scenario where the computational budget is limited, possibly due to the multitude of scenarios to evaluate or a time restriction, evaluating all the trajectories is not always a viable option. 
Having a method to determine, in a robust statistical manner, where the aircraft are most likely to be, is the first step in efficiently identifying high-risk launch sites.


Focussing on the work done in the aerospace sector, there have already been great advances in automatically clustering of aircraft trajectories on a common flightpath.
One clustering method re-samples the trajectories to fit in a vector of fixed size, after which the size of the vector is reduced using \gls{pca}. The data are then clustered using \gls{dbs} \cite{gariel2011trajectory}, or \emph{k-means} clustering \cite{Eckstein2009}. Here the \gls{dbs} \cite{Ester1996} allows for the filtering of outliers, resulting in a more robust clustering method compared to using \emph{k-means} clustering. Another interesting method used to cluster the aircraft trajectories is based on Fourier coefficients \cite{Annoni2012}. The major difference here is that the trajectories are not merely re-sampled, but represented as Fourier-coefficients that effectively parametrises the aircraft trajectories. However, it should be said that the parametrisation is limited here to two spatial dimensions, ignoring the vertical variation in the flight-paths. 


Automatic clustering (i.e. discovery of flight-paths) is an important first step, however, it holds little information about the level of dispersion of the trajectories within a given cluster, i.e. around the nominal flight-path. A method has been proposed by Salaun et al.\cite{salaun2012} to calculate the probabilities by re-sampling, and fitting a univariate Gaussian in both the lateral and vertical direction perpendicular to the mean trajectory. Using this approach, they were successful in creating a tunnel through which a percentage of aircraft trajectories manoeuvre. More recently, a similar approach of modelling trajectories as \acrfull{gps} has been developed by the authors\cite{eerland2015}. This allows the aircraft trajectories to be treated in a continuous manner and also models the covariance between the lateral and vertical direction.


In this paper, we present a step-by-step method to replace a large data-set of historical trajectories, with a number of representative trajectories that have none-the-less captured the dispersion in the original large data-set. The trajectory data are clustered, after which a probabilistic model is generated for each individual cluster. This probabilistic model is then used to generate weighted representative trajectories, each capturing a fraction of the whole cluster. The aim here is to reduce the computational cost of calculations that are done in sub-sequential steps. Such calculations can be either focussed on calculating noise footprints, or be aimed at performing a strategic analysis with the focus on safeguarding the airspace infrastructure. The latter is evaluated to demonstrate the effectiveness of the proposed method when generating a footprint on ground level.

This paper is organised as follows. In \cref{sec:techniques}, each step is explained, resulting in a guide how to transform the original trajectory data into a set of weighted representative trajectories.
Next, in \cref{sec:results}, the methods are applied in a case study. This includes in an evaluation to compare the original trajectory data with a set of representative trajectories. The set of representative trajectories are generated in two ways, one takes into account the dispersion in lateral direction only, while the other also includes the vertical dispersion.
Finally, in \cref{sec:conclusion}, we conclude the paper with remarks and recommendations for future work.

\section{Methods} \label{sec:techniques}
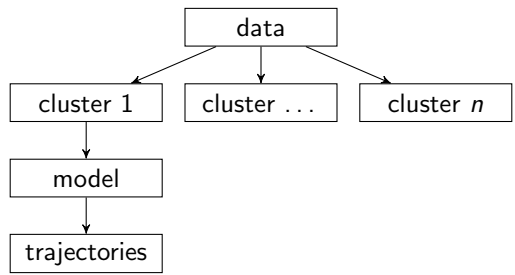
\begin{figure}
\centering
\begin{tikzpicture}[->,>=stealth',level/.style={sibling distance = 2.3cm/#1,
  level distance = 1cm}]
\node [arn_x] {data}
    child{ node [arn_x] {cluster 1}
            child{ node [arn_x, distance=5cm] {model}
            	child{ node [arn_x, distance=5cm] {trajectories}
				}
            }
    	}
    child{ node [arn_x] {cluster \ldots}
    	}
    child{ node [arn_x] {cluster \emph{n}}
    	}
;
\end{tikzpicture}
\caption{Procedure to obtain representative trajectories based on data.}
\label{fig:procedure}
\end{figure}

This section reviews the techniques used to replace historical data with representative trajectories. 
The first part focuses on clustering the trajectory data by identifying groups of trajectories with similar flight-path.
The second part focuses on estimating the dispersion of each individual cluster along the flight-path.
The third and final part is aimed at creating weighted trajectories that represent the aircraft trajectories flying through the airspace.
The complete procedure is seen in \cref{fig:procedure}, where the data are first clustered, then modelled and finally representative trajectories are generated (leaving the possibility of selection).

\subsection{Clustering the Trajectories} \label{sec:tech-clustering}
In order to cluster the trajectories based on a common flightpath, it is assumed that in the reference frame, the location of the runways are known and there is a human-in-the-loop, i.e. the process is not fully automated. The clustering technique takes the shape of the following three-step approach:
\begin{enumerate}
\item clustering trajectories as \emph{approach} or \emph{departure}
\item clustering the trajectories by run-way
\item re-sampling of trajectories
\item dimension reduction with \gls{pca}
\item \gls{dbs} clustering
\end{enumerate}
The first step distinguishes the trajectory data between approach and departure. In the case where these meta-data are not included, the location of the airport is sufficient to identify whether a trajectory either ends at the airport (approach), or is leaving (departure). E.g. if the Euclidean distance between the airport and the first point of the trajectory is smaller than the Euclidean distance between the airport and the last point of the trajectory, it is likely to be a departure.
The second step requires information about the location of the runway in a similar reference frame as the trajectories. While it reduces the generality, it does allow for a clean separation per runway.
%

Steps $3-5$ are mostly similar to the procedure as presented in Gariel et al.\cite{gariel2011trajectory}.
In step $3$ the trajectories are re-sampled to a vector of a fixed size using uniform spacing based on the index number. Note that re-sampling over $30$ steps result in a $[1 \times 90]$ vector as aircraft trajectories have $3$ dimensions, and the dimensions are concatenated. Furthermore, the re-sampling is done per individual trajectory, while the next step, the reduction of the vector size, considers the entire data-set.
Next, in step $4$, these vectors are reduced in size using \gls{pca}. Here the principle components with the largest variance are kept, while the components with little variation are ignored. In this step it is assumed that the components with the largest variation are most important for the clustering.
The clustering occurs in the final step, where the trajectories are clustered using \gls{dbs}.
Here, the $\epsilon$-neighbourhood parameter \gls{eps} of the \gls{dbs} algorithm needs to be set on a case-by-case basis, a smaller \gls{eps} will result in more clusters with fewer trajectories, whereas with a larger \gls{eps} results in less clusters with more trajectories.
In the final step all clusters with less than a user-specified number of trajectories will be ignored, effectively removing the outliers.

\subsection{Modelling the Spatial Distribution} \label{sec:tech-modelling}
In this section a short overview of the modelling technique is provided. For a complete description, please see Eerland and Box\cite{eerland2015}.
Essentially, there are two steps in modelling the spatial distribution:
\begin{itemize}
    \item normalising the trajectory data (per cluster, see previous section)
    \item learn model parameters via maximum likelihood estimation
\end{itemize}

The first step assures that the dimensions are of the same scale. In particular for aircraft trajectories this is an important step, as usually the distance covered horizontally is much larger than the distance travelled vertically. When estimating the parameters, this can lead to computational problems, as such, the trajectory data are normalised such that each dimension fits on a $[0,1]$ range. This transformation can be reverted once the model has been created.
Furthermore, normalised time $\gls{t}$ is introduced to align all points of the trajectories at the start at the end. The start is set to be $\gls{t} = 0$ and the end is $\gls{t} = 1$, where the points in between are set proportionally. E.g. if the first point occurs at $0$ seconds and the last point occurs at $20$ seconds, the point at $12$ seconds will have the normalised time $\gls{t} = 12/20 = 0.6$.
As such, each point in the trajectory is described as a $3 \times 1$ vector $\gls{yv}(\gls{t})$, holding eastings, northings and altitude at the normalised time $\gls{t}$.

In the second step the model parameters are estimated. Here the parameters consist of the mean function $\gls{m}(\gls{t})$, covariance kernel $\gls{k}(\gls{t},\gls{t}')$ and noise precision term $\gls{beta}$. These parameters capture the underlying function $\gls{yv}(\gls{t})$ according to the following relation:
\begin{equation} \label{eq:gp-y}
  \gls{yv}(\gls{t}) \sim \mathcal{GP}(\gls{m}(\gls{t}), \gls{k}(\gls{t},\gls{t}\sp{\prime}))
\end{equation}
where
\begin{equation} \label{eq:expectation-y}
  \gls{m}(\gls{t}) = \gls{E} [ \gls{yv} ] = \gls{H}(\gls{t})  \gls{mu}
\end{equation}
\begin{equation} \label{eq:covariance-y}
  \gls{k}(\gls{t},\gls{t}\sp{\prime}) = \gls{E} [ (\gls{yv}(\gls{t}) - \gls{m}(\gls{t})) (\gls{yv}(\gls{t}\sp{\prime}) - \gls{m}(\gls{t}\sp{\prime})) ] = \gls{H}(\gls{t}) \gls{sigma} \gls{H}^{\intercal}(\gls{t}\sp{\prime}) + \gls{beta}^{-1} \gls{I}
\end{equation}
In these equations the mean function $\gls{m}(\gls{t})$ and the covariance kernel $\gls{k}(\gls{t},\gls{t}')$ are captured using $J$ basis functions. Here the discrete number of parameters captured in the $3 J \times 1$ vector $\gls{mu}$ and $3  J \times 3 J$ matrix $\gls{sigma}$, are converted from the discrete domain to the continuous domain via the block-diagonal $3 \times J$ matrix $\gls{hv}(\gls{t})$. The basis functions $\gls{hv}(\gls{t})$ consists of $3$ blocks, corresponding with the number of dimensions found in aircraft trajectories.

Next, for the estimation the \gls{em} algorithm is applied, this deals with the chicken-egg paradigm. More specifically, $\gls{beta}$ is needed to estimate $\gls{m}(\gls{t})$ and $\gls{k}(\gls{t},\gls{t}')$, and $\gls{m}(\gls{t})$ and $\gls{k}(\gls{t},\gls{t}')$ are needed to estimate $\gls{beta}$. Basically, each individual trajectory is captured in the model described by the couple $\gls{m}(\gls{t})$ and $\gls{k}(\gls{t},\gls{t}')$, however not perfectly, thus the remaining error is captured in $\gls{beta}$. And by doing so, maximizing the likelihood of these three terms using \gls{em}, prevents the probabilistic model to over-fit on the data (assuming the remaining error is Gaussian distributed).

Initially $\gls{mu}$ (a $3 J \times 1$ vector) is assumed $\mathbf{0}$, and $\gls{sigma}$ (a $3  J \times 3 J$ matrix) is assumed $a \gls{I}$, where $\gls{I}$ is the identity matrix and $a$ an arbitrarily large number. This represents a \emph{not very informative} prior and reflects the concept that no initial knowledge is available. The noise precision term $\gls{beta}$ can be set high (e.g. in the order of magnitude of ${10}^{3}$), to reflect that the measurements are exact and the remaining error low.

Using these initial parameters, the expected parameter values $\gls{wv}$ (the E-step of the \gls{em} algorithm) are:
\begin{equation} \label{eq:Ewn}
  \gls{E} [ \gls{wv}_n ] = \gls{sigma1}_n ( \gls{beta} \gls{H}_n^\intercal \gls{yv}_n + \gls{sigma}^{-1} \gls{mu} )
\end{equation}
\begin{equation} \label{eq:Ewnwn}
  \gls{E} [ \gls{wv}_n \gls{wv}_n^\intercal ] = \gls{sigma1}_n + \gls{E} [ \gls{wv}_n ] \gls{E} [ \gls{wv}_n^\intercal ]
\end{equation}
where
\begin{equation} \label{eq:sn}
  \gls{sigma1}_n^{-1} = \gls{sigma}^{-1} + \gls{beta} \gls{H}_n^\intercal \gls{H}_n
\end{equation}
In these equations $n$ represents the individual trajectory, and $N$ equates the total number of trajectories found in the cluster.

Next, the likelihood is maximized with respect to the model parameters (the M-step of the \gls{em} algorithm) using:
\begin{equation} \label{eq:muh}
  \gls{muh} = \frac{1}{N} \sum_{n=1}^N \{ \gls{E} [ \gls{wv}_n ] \}
\end{equation}
\begin{equation} \label{eq:sigmah}
  \gls{sigmah} = \frac{1}{N} \sum_{n=1}^N \{ \gls{E} [ \gls{wv}_n \gls{wv}_n^\intercal ] - 2 \gls{E} [ \gls{wv}_n^\intercal ] \gls{mu} + \gls{mu} \gls{mu}^\intercal \}
\end{equation}
\begin{equation} \label{eq:betah}
  \frac{1}{\gls{betah}} = \frac{1}{ 3 M^* } \sum_{n=1}^N \{ \gls{yv}_n^\intercal \gls{yv}_n - 2 \gls{yv}_n^\intercal ( \gls{H}_n \gls{E} [ \gls{wv}_n ] ) + \Tr ( \gls{H}_n^\intercal \gls{H}_n \gls{E} [ \gls{wv}_n \gls{wv}_n^\intercal ] ) \}
\end{equation}
where the hat seen in $\gls{muh}$, $\gls{sigmah}$ and $\gls{betah}$ signifies an approximation. These two steps in the \gls{em} algorithm are repeated until the likelihood is converged, where the negative log-likelihood itself can be evaluated using:
\begin{multline} \label{eq:log-likelihood}
  - \ln \gls{L} =
  - \frac{3 M^*}{2} \ln (\gls{beta})
  + \frac{\gls{beta}}{2} \sum_{n=1}^N \{ \gls{yv}_n^\intercal \gls{yv}_n - 2 \gls{yv}_n^\intercal ( \gls{H}_n \gls{wv}_n ) + \Tr ( \gls{H}_n^\intercal \gls{H}_n \gls{wv}_n \gls{wv}_n^\intercal ) \} \\
  + \frac{N}{2} \ln ( | \gls{sigma} | )
  + \frac{1}{2} \sum_{n=1}^N \{ \Tr( \gls{sigma}^{-1} ( \gls{wv}_n \gls{wv}_n^\intercal - 2 \gls{wv}_n^\intercal \gls{mu} + \gls{mu} \gls{mu}^\intercal )  ) \}
\end{multline}
where
\begin{equation} \label{eq:m-star}
  M^* = \sum_{n=1}^N \{ M_n \}
\end{equation}
and $M_n$ represents the total number of points in $\gls{yv}_n$, thus $M^*$ embodies the total number of points in the entire cluster.

The difference between two sequential log-likelihood evaluations is used as a stopping criteria, at this point the model approximation is assumed sufficient. Due to the nature of the \gls{em} algorithm, it will always be considered an approximation. 

The approximated parameters can now be interested in the model, seen in \cref{eq:gp-y}, to estimate the probabilistic model at any $\gls{t}$ in the domain $\gls{t} = [0,1]$. This model allows itself to be expressed in a multivariate Gaussian distribution as a function of $\gls{t}$, which will be used to generate \emph{weighted} representative trajectories in the next section.

\subsection{Generating representative trajectories} \label{sec:tech-selection}
The previous section described how to estimate the probabilistic model. This section provides a method to convert this model to \emph{weighted} trajectories that represent the entire cluster.

For a $3$-dimensional vector $\gls{yv}(\gls{t})$, the multivariate Gaussian distribution takes the form:
\begin{equation}
\gls{N}(\gls{m}, \gls{k}) = \frac{1}{\sqrt{(2 \pi)^{3/2}} } \frac{1}{ {| \gls{k} |}^{1/2} }   \exp \left\{ - \frac{1}{2} \text{MD}(\gls{yv},\gls{m},\gls{k}) \right\}
\end{equation}
where $\text{MD}$ represents the Mahalanobis distance:
\begin{equation}
\text{MD}(\gls{yv},\gls{m},\gls{k})) = (\gls{yv} - \gls{m})^\intercal \gls{k}^{-1} (\gls{yv} - \gls{m})
\end{equation}
And note that the dependence on $\gls{t}$ has been dropped for readability.

Furthermore, at a constant Mahalanobis distance, this equation takes the form of an ellipsoid described by ${| \gls{k} |}^{-1/2}$, centred at $\gls{m}$.
In this scenario the axes of the covariance ellipse are given by the eigenvectors $\mathbf{r}$ of the covariance kernel $\gls{k}$.
The corresponding lengths, for an ellipse with unit Mahalanobis radius, are given by the square roots of the corresponding eigenvalues $\lambda$.
Both can be found using the eigenvalue decomposition of the matrix $\gls{k}$.
\begin{equation}
    \mathbf{R} \boldsymbol{\Lambda} \mathbf{R}^{-1} = \gls{k}
\end{equation}
where
\begin{equation}
    \mathbf{R} = \left[ \mathbf{r}_1 ~ \mathbf{r}_2 ~ \mathbf{r}_3 \right]
\end{equation}
and
\begin{equation}
    \boldsymbol{\Lambda} =
    \begin{bmatrix}
    \lambda_1 & 0 & 0 \\
    0 & \lambda_2 &  0 \\
    0 & 0 &  \lambda_3
    \end{bmatrix}
\end{equation}
In short, the shape of the ellipsoid is described by the eigenvalues found in $\boldsymbol{\Lambda}$, where $\gls{m}$ and $\mathbf{R}$ are merely a translation and rotation respectively.
The equation for the ellipsoid is given by:
\begin{equation}
    \frac{\gls{y}_1^2}{\lambda_1} + \frac{\gls{y}_2^2}{\lambda_2} + \frac{\gls{y}_3^2}{\lambda_3} = 1
\end{equation}

The plane perpendicular to the mean function at time $\gls{t}$, can be described with a unit normal vector $\mathbf{n}$.
As $\gls{m}(\gls{t})$ is continuous, this unit normal vector $\mathbf{n}$ can be both derived, or calculated numerically:
\begin{equation}
    \mathbf{n} = \frac{ \gls{m}(\gls{t}+d\gls{t}/2) - \gls{m}(\gls{t}-d\gls{t}/2) }{ d\gls{t} }
\end{equation}
where $d\gls{t}$ is an arbitrarily small number.
The mathematics required to calculate the intersection between the plane and ellipsoid is given in Klein \cite{klein2012ellipsoid}. By doing so, the \emph{two-dimensional} ellipse can be evaluated at any angle.
However, it's important to note that the plane generated at time $\gls{t}$ intersects multiple ellipsoids, it's therefore necessary to evaluate all those that intersect and store the point corresponding with the largest deviation from the centre point (provided by $\gls{m}(\gls{t})$).

\begin{figure}
 \begin{subfigmatrix}{2}
  \subfigure[Cross-section of the representative trajectories (indicated by blue dots), when only taking the lateral dispersion into account. At the top is a schematic representation of the Gaussian distribution, placed as a reference.]
  {
    \centering
    \begin{tikzpicture}
    \begin{axis}[
        no markers, domain=-2.5:2.5, samples=100,
        axis lines = none,
        every axis x label/.style={at=(current axis.right of origin),anchor=west},
        height=4cm, width=6.7cm,
        xtick={-2,-1,0,1,2}, ytick=\empty,
        enlargelimits=false, clip=false, axis on top,
        grid = major
        ]
        \addplot [fill=black!60, draw=none, domain=-2.5:2.5] {gauss(0,1)} \closedcycle;
        \addplot [fill=black!40, draw=none, domain=-1.5:1.5] {gauss(0,1)} \closedcycle;
        \addplot [fill=black!20, draw=none, domain=-0.5:0.5] {gauss(0,1)} \closedcycle;
        \addplot [very thick,black] {gauss(0,1)};

    \end{axis}

    \begin{scope}[shift={(2.55,-1)}]
        \draw[fill=black!60] (-2.5,-.5) -- (-2.5,.5) -- (2.5,.5) -- (2.5,-.5) -- cycle;
        \draw[fill=black!40] (-1.5,-.5) -- (-1.5,.5) -- (1.5,.5) -- (1.5,-.5) -- cycle;
        \draw[fill=black!20] (-0.5,-.5) -- (-0.5,.5) -- (0.5,.5) -- (0.5,-.5) -- cycle;
        \draw[fill=blue] (-2,0) circle (.1cm);
        \draw[fill=blue] (-1,0) circle (.1cm);
        \draw[fill=blue] (0,0) circle (.1cm);
        \draw[fill=blue] (1,0) circle (.1cm);
        \draw[fill=blue] (2,0) circle (.1cm);
        \node at (-2,.75) {2};
        \node at (-1,.75) {1};
        \node at (0,.75) {0};
        \node at (1,.75) {1};
        \node at (2,.75) {2};
    \end{scope}
    \end{tikzpicture}
    \label{fig:cross-section-flat}
  }
  \subfigure[Cross-section of the representative trajectories (indicated by blue dots) when taking both the lateral and vertical dispersion into account.]
  {
    \centering
    \begin{tikzpicture}
    \draw[fill=black!60] (0,0) circle (2.5cm);
    \draw[fill=black!40] (0,0) circle (1.5cm);
    \foreach \theta in {22.5,67.5,...,382.5}
    {
        \draw (0,0) -- (\theta:2.5cm);
    }
    \draw[fill=black!20] (0,0) circle (.5cm);   

    \foreach \dist in {0cm, 1cm, 2cm}       
    {
        \foreach \theta in {0,45,...,360}
        {
            \draw[fill=blue] (\theta:\dist) circle (.1cm);
        }
    }
    \end{tikzpicture}
    \label{fig:cross-section-round}
  }
 \end{subfigmatrix}
 \caption{Two options to obtain representative trajectories.}
 \label{fig:cross-section}
\end{figure}
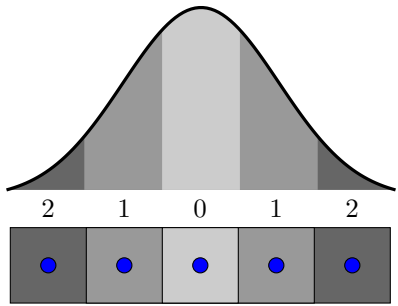
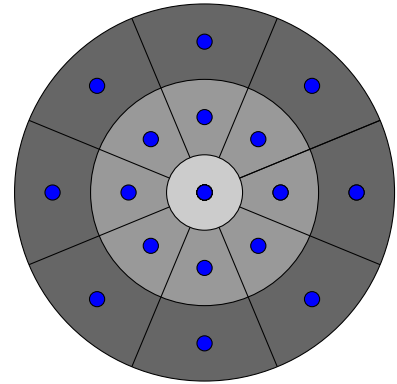

To obtain a representative trajectory, the ellipse at any specific angle, which is a single point in a three-dimensional space, can be evaluated over $\gls{t} = [0,1]$ in any number of steps. In this paper $100$ steps are used. Thus the combination of a constant Mahalanobis distance (representing a confidence interval, to be discussed next) and a given angle provides one representative trajectory.
However, when it comes to selecting the angles to obtain a selection of representative trajectories, there are an infinite number of options. In this paper two options are compared.
In the handbook on generating noise contours \cite{ECAC2005}, only the dispersion in lateral direction is taken into account. Here the cross-section containing the artificial trajectories appears like \cref{fig:cross-section-flat}, where the Gaussian distribution is included as a reference. In this figure, $5$ trajectories are shown to capture a given percentage (the area under the curve).
This corresponds with the \gls{cdf}, which is equal to chi-square with $1$ degree of freedom. The resulting weight per trajectory is shown in \cref{tab:weight-flat}.  
The area under the curve described by the Gaussian corresponds with a confidence interval (thus capturing a certain percentage of the complete data) - and this percentage is divided over multiple trajectories due to symmetry.
E.g. for the case of lateral dispersion only, in the range $[0.5^2, 1.5^2]$, the total percentage $48.3461 \%$ is divided over two trajectories centred at a standard deviation of one ($N = 2$), resulting in $24.1730 \%$ per trajectory.
When including the vertical dispersion, the model show more similarity to \cref{fig:cross-section-round}. As it now encompasses two dimensions, the chi-square with $2$ degrees of freedom is used. Combined with angles at various ranges, there are $17$ representative trajectories. The weight per trajectory is shown in \cref{tab:weight-round}.
In the evaluations seen further on in this paper, these percentages will be multiplied with the total traffic to obtain representative number of trajectories. I.e. the percentages shown here, are the weights used in the calculations.

\begin{table}
    \centering
    \caption{Weights generated using the chi-square distribution table ($1$ degree of freedom).} \label{tab:weight-flat}
    \begin{tabular}{ | c | c | c | c | c | }
     \hline
    \textbf{range} & \textbf{total percentage} & \textbf{percentage per} \\
     & \textbf{captured} & \textbf{trajectory} \\
      \hline
      $[0.0^2, 0.5^2]$ & $38.29 \%$ & $38.29 \%$  \textit{(N = 1)} \\
      $[0.5^2, 1.5^2]$ & $48.35 \%$ & $24.17 \%$ \textit{(N = 2)} \\
      $[1.5^2, 2.5^2]$ & $12.12 \%$ & $6.06 \%$ \textit{(N = 2)} \\
      $[0.0^2, 2.5^2]$ & $98.76 \%$ & $98.76 \%$ \textit{(N = 5)} \\
      \hline
    \end{tabular}
\end{table}

\begin{table}
    \centering
    \caption{Weights generated using the chi-square distribution table ($2$ degrees of freedom).} \label{tab:weight-round}
    \begin{tabular}{ | c | c | c | c | c | }
     \hline
    \textbf{range} & \textbf{total percentage} & \textbf{percentage per} \\
     & \textbf{captured} & \textbf{trajectory} \\
      \hline
      $[0.0^2, 0.5^2]$ & $11.75 \%$ & $11.75 \%$ \textit{(N = 1)} \\
      $[0.5^2, 1.5^2]$ & $55.78 \%$ & $6.97 \%$ \textit{(N = 8)} \\
      $[1.5^2, 2.5^2]$ & $28.07 \%$ & $3.51 \%$ \textit{(N = 8)} \\
      $[0.0^2, 2.5^2]$ & $95.61 \%$ & $95.61 \%$ \textit{(N = 17)} \\
      \hline
    \end{tabular}
\end{table}

\pgfmathdeclarefunction{gauss}{2}{%
  \pgfmathparse{1/(#2*sqrt(2*pi))*exp(-((x-#1)^2)/(2*#2^2))}%
  }

\section{Results} \label{sec:results}
This \namecref{sec:results} applies the step-by-step guide as described in \cref{sec:techniques} on a case-study. The case-study here is based on aircraft trajectories near an airport as measured by ground radar.
The first step is to cluster the trajectories, after which the individual clusters of trajectories (sorted per common flight-path) are modelled, and in the final step a sub-set of representative trajectories is generated.
For the evaluation a $100 \times 100$ grid is placed near the airport at a ground level. For each of these grid-points the percentage of the total number of trajectories in the (post-clustered) data that come within $300$ metres (as measured from the centre) is calculated. Under the assumption that each trajectory is equally likely to occur, it can be said that a person standing in the area with the highest percentage is more likely to `spot' an aircraft within a $300$ metres radius. Where in the areas with $0\%$ there is no chance at all.
This relates back to the introduction where it was pointed out that there is evidence that terrorists have access to \gls{rpgs}, and that these weapons are effective up to a range of $300$ metres against a moving target, in this case an aircraft.

\begin{figure}
\centering
    \includegraphics[width=\singlepic]{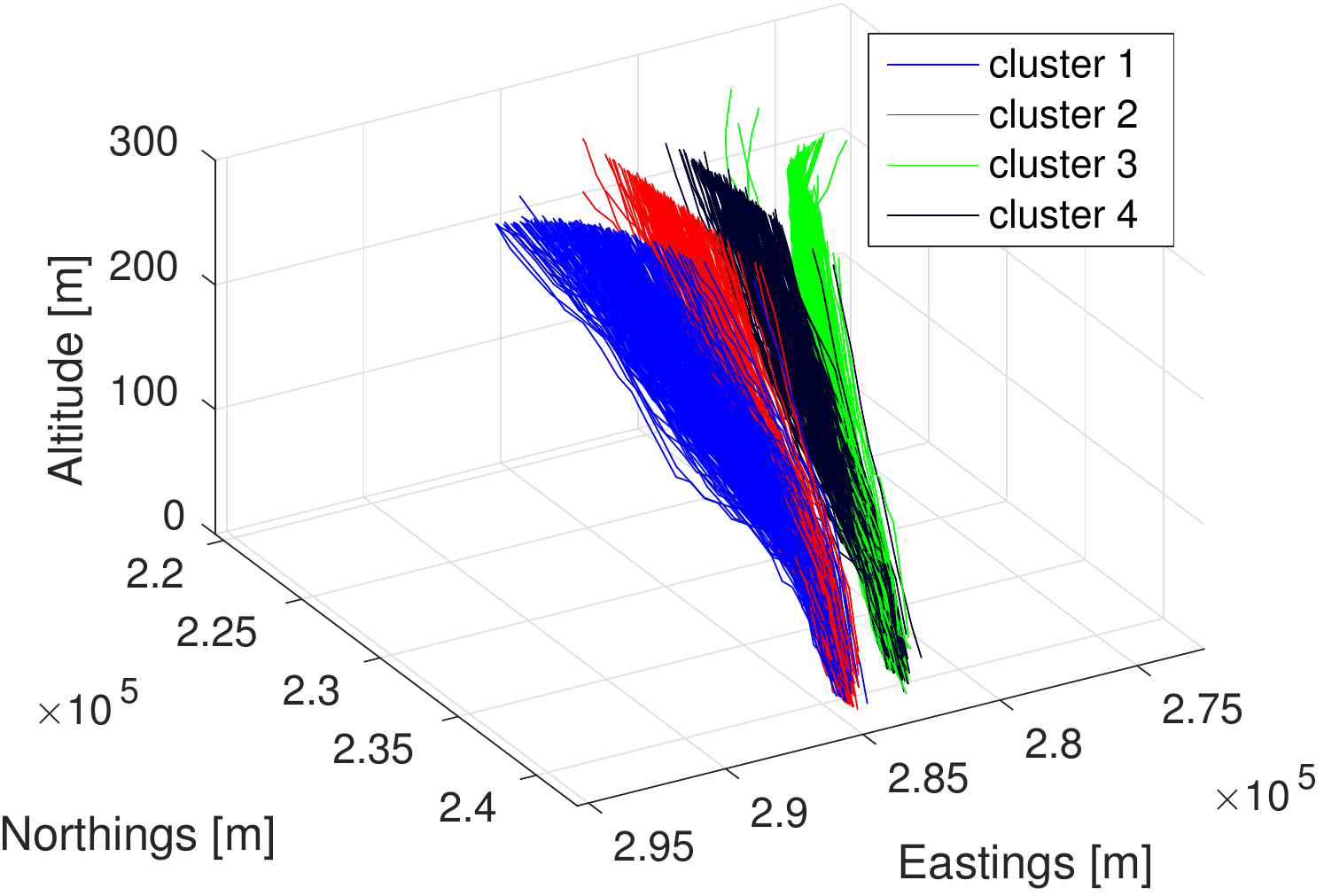}
    \caption{Three-dimensional view of all the clustered original trajectory data.}
    \label{fig:clustered-xyz}
\end{figure}

The trajectory data are clustered according to the technique outlined in \cref{sec:tech-clustering}, where the result is visible in \cref{fig:clustered-xyz}. For the analysis in this section, only the \emph{departure} trajectories are included to keep the amount of trajectories manageable and the eventual evaluation of a reasonable scale. The number of trajectories found in each cluster is shown in \cref{tab:number-of-trajectories}.
The $\epsilon$-neighbourhood parameter \gls{eps} of the \gls{dbs} algorithm is set to $0.20$ and the minimum number of trajectories in each cluster is $25$, causing just over $19\%$ (193 trajectories) to be considered outliers. These outliers are not shown.
Furthermore, as the ultimate goal is to create a footprint based on the distance of $300$ metres, only the parts of the trajectories below an altitude of $500$ metres are used in the analysis. 

\begin{table}
    \centering
    \caption{Number of trajectories found in each cluster.} \label{tab:number-of-trajectories}
    \begin{tabular}{ | c | c | }
     \hline
    \textbf{cluster} & \textbf{number of trajectories} \\
      \hline
      $1$ & $267$ \\
      $2$ & $194$ \\
      $3$ & $178$ \\
      $4$ & $176$ \\
      \hline
    \end{tabular}
\end{table}

\begin{figure}
\centering
    \includegraphics[width=\singlepic]{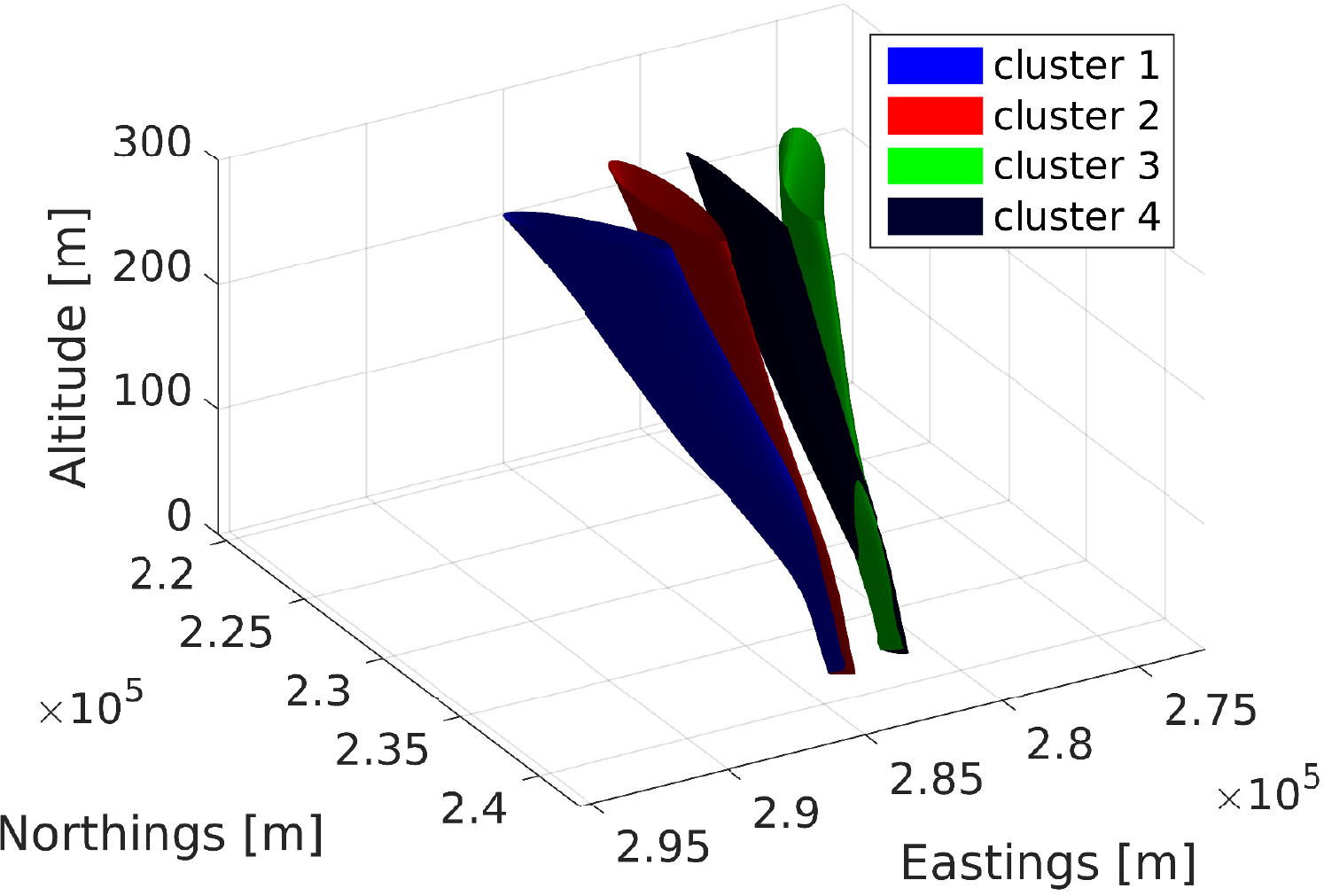}
    \caption{Three-dimensional view of all models - where the volume corresponds with a standard deviation of two.}
    \label{fig:model-xyz}
\end{figure}

Based on the method described in \cref{sec:tech-modelling}, a model is generated for each cluster.
The basis functions $\gls{hv}(\gls{t})$ in this paper consist of $17$ radial basis functions, uniformly distributed over the interval $\gls{t} = [0,1]$. And when including the bias function, the total number of basis function $J$ is $18$. The convergence criteria is set such that the difference between two sequential iterations, and evaluation of the log-likelihood, is at most $10^{-3}$.
The result is shown in \cref{fig:model-xyz}, where the volume represents the range corresponding with a standard deviation of two ($\gls{std} = 2$).


\begin{figure}
\centering
    \includegraphics[width=\singlepic]{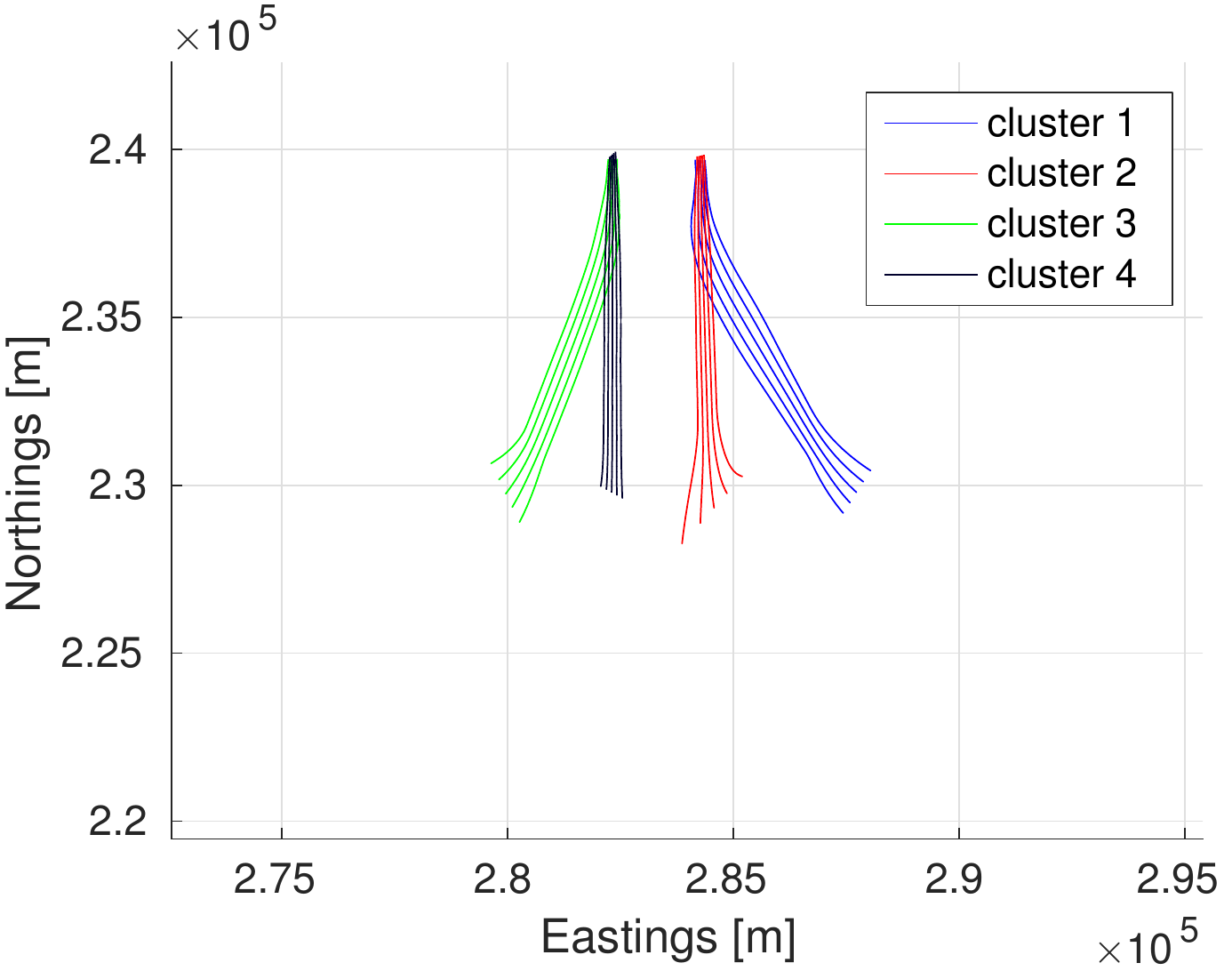}
    \caption{Top-view of the representative trajectories (the lateral direction only).}
    \label{fig:artificial-5-xy}
\end{figure}

\begin{figure}
\centering
    \includegraphics[width=\singlepic]{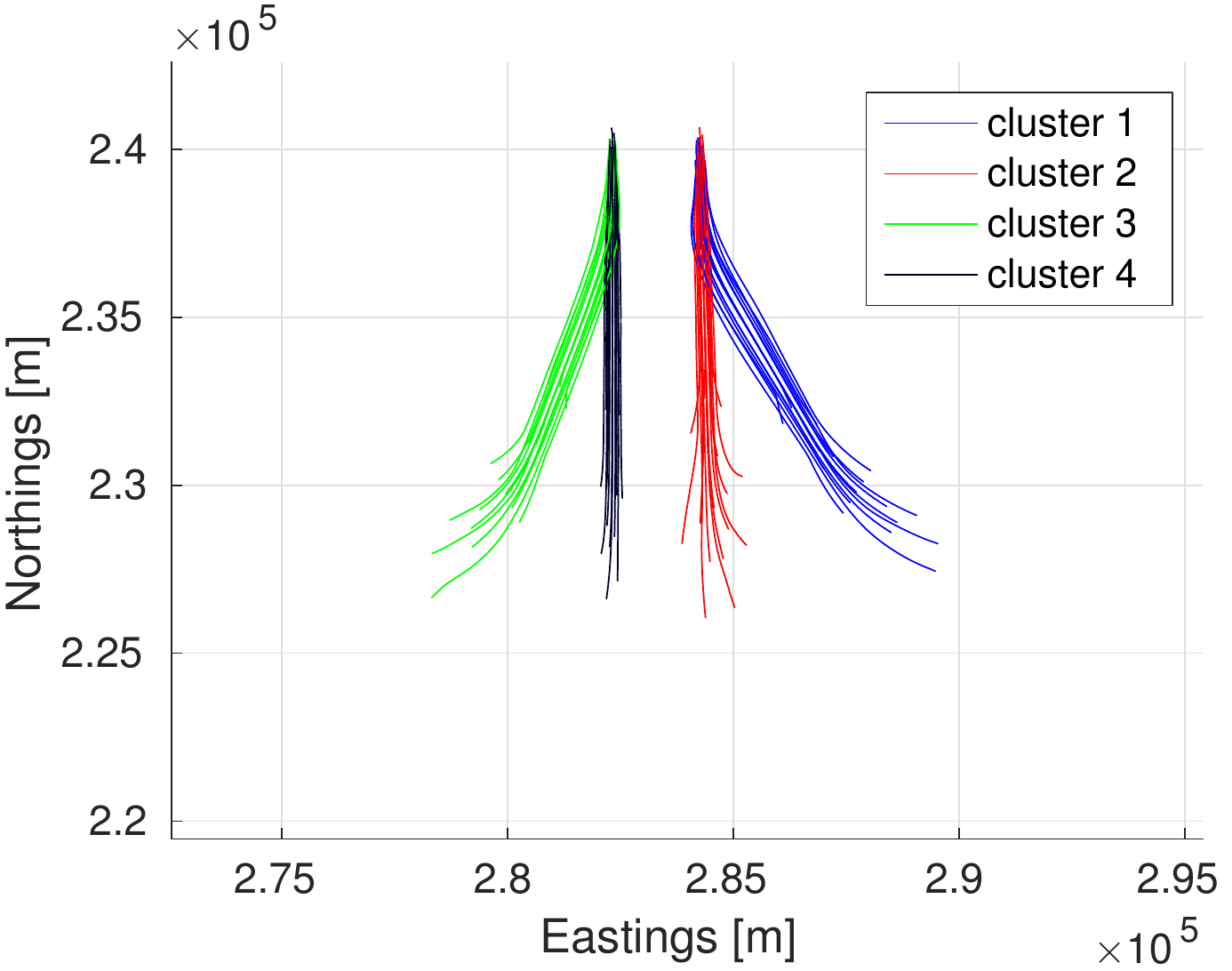}
    \caption{Top-view of the representative trajectories (lateral and vertical direction).}
    \label{fig:artificial-17-xy}
\end{figure}

In \cref{sec:tech-selection} two approaches to generate representative trajectories were described.
One automatically generates the trajectories while only taking into account the lateral dispersion. The cross-section of this `flat' version is seen in \cref{fig:cross-section-flat}, and the resulting top-view is visible in \cref{fig:artificial-5-xy}. This modelling approach is currently being used in calculating noise footprints around airports, as described in the official publication \cite{ECAC2005}.
The other approach takes into account both the lateral and vertical dispersion simultaneously. The cross-section of this `round' version is seen in \cref{fig:cross-section-round}, and the resulting top-view is visible in \cref{fig:artificial-17-xy}.

\begin{figure}
\centering
    \includegraphics[width=\singlepic]{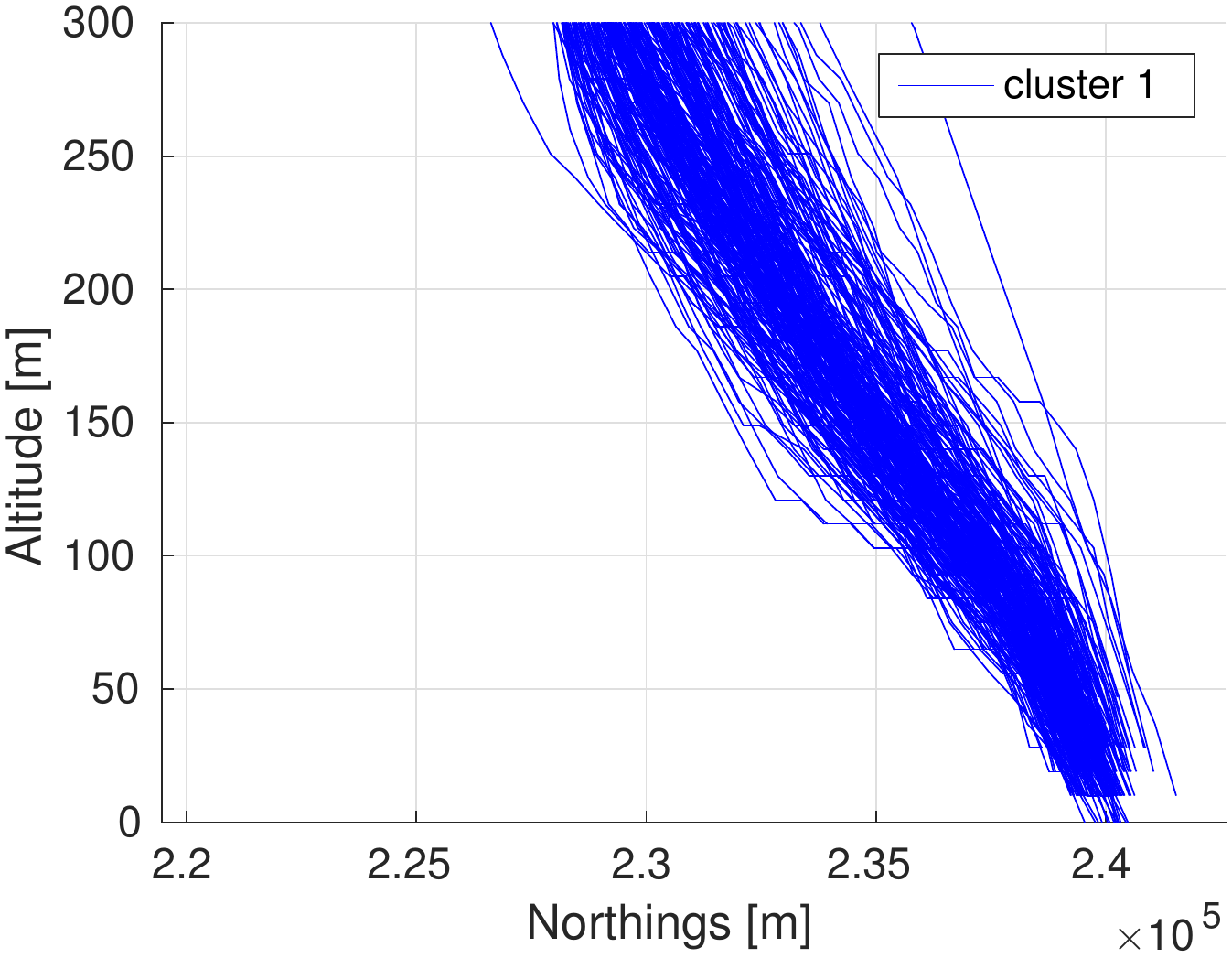}
    \caption{Side-view of the largest cluster of the original trajectory data.}
    \label{fig:lclustered-yz}
\end{figure}

\begin{figure}
\centering
    \includegraphics[width=\singlepic]{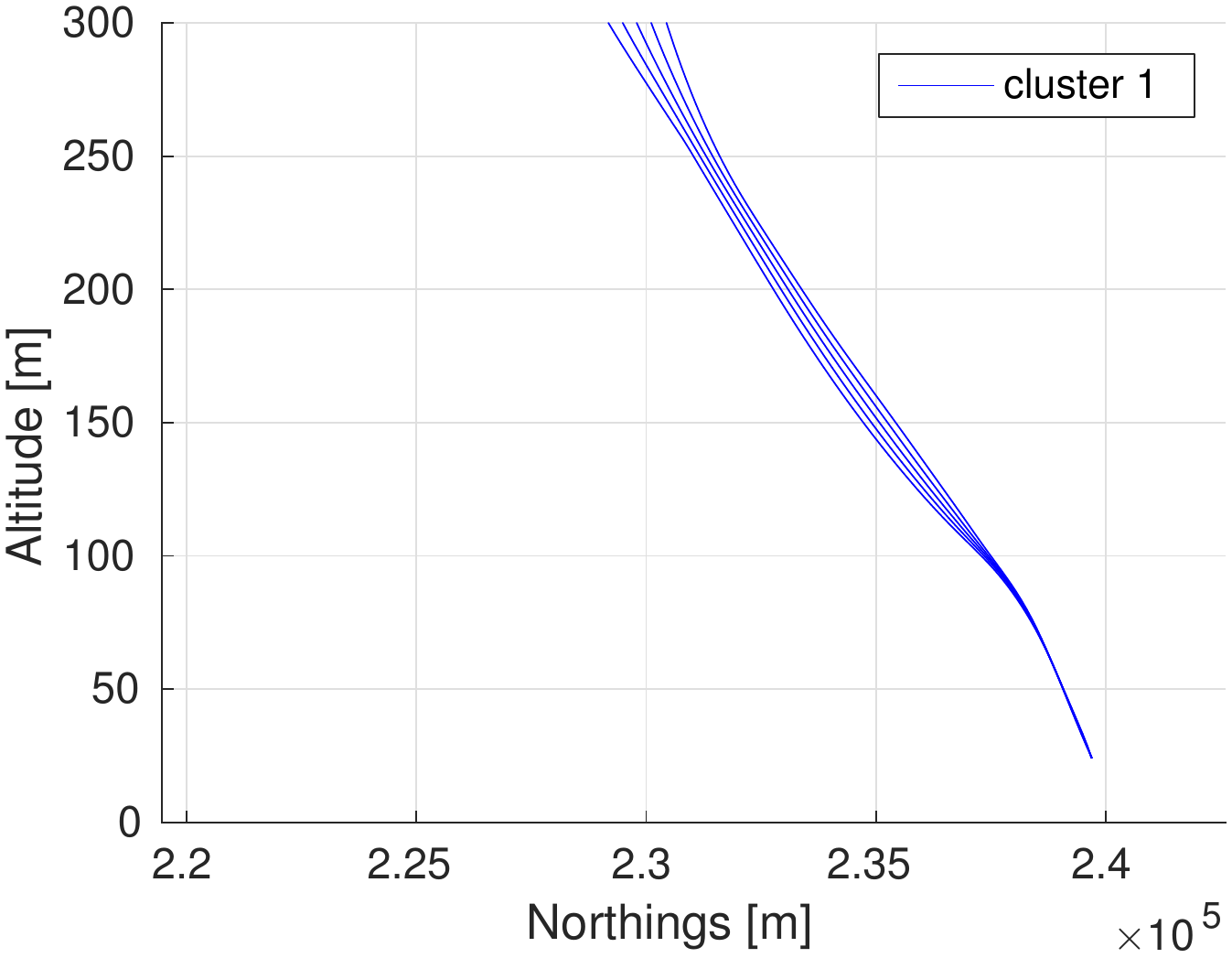}
    \caption{Side-view of the representative trajectories (lateral direction only).}
    \label{fig:lartificial-5-yz}
\end{figure}

\begin{figure}
\centering
    \includegraphics[width=\singlepic]{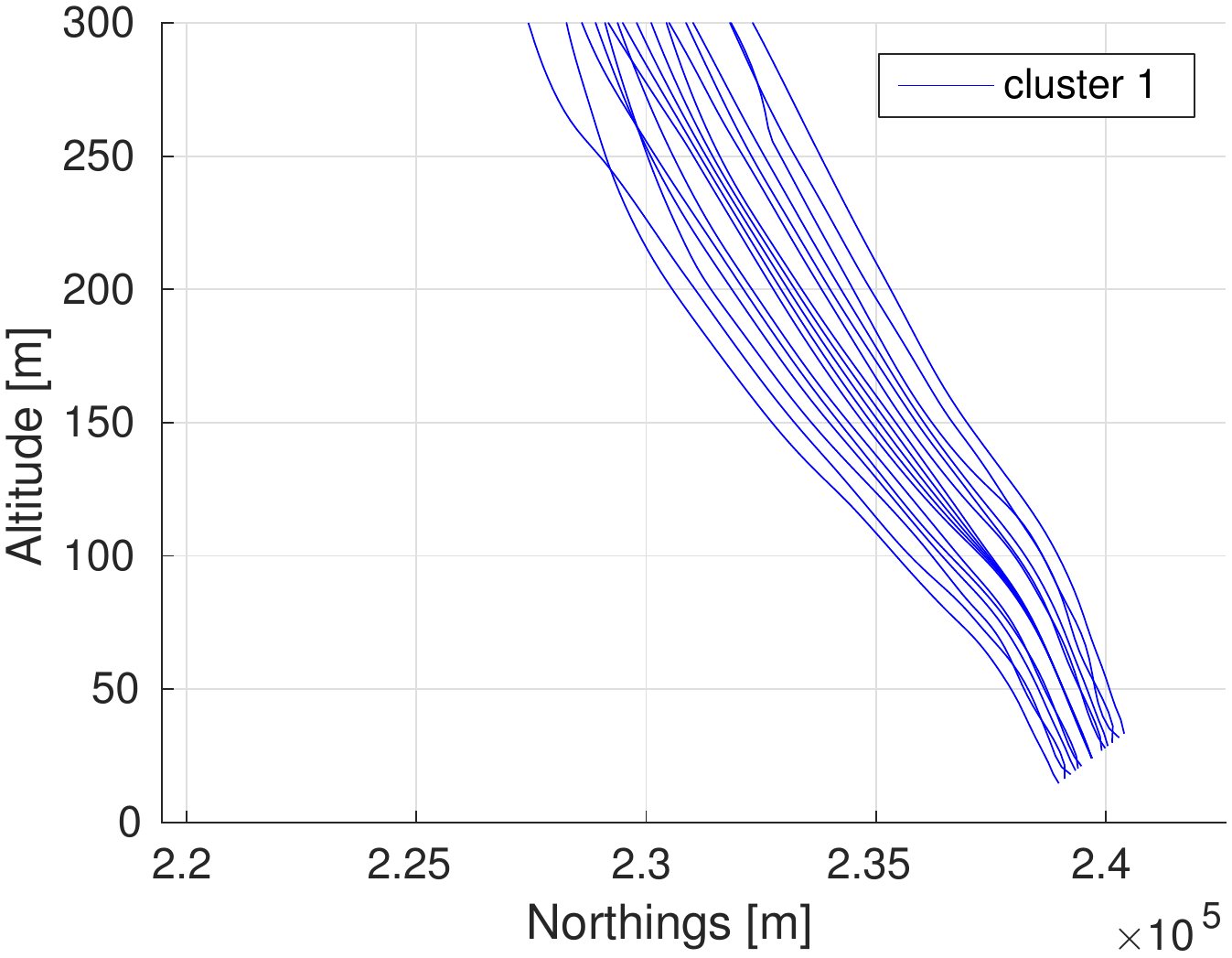}
    \caption{Side-view of the representative trajectories (lateral and vertical direction).}
    \label{fig:lartificial-17-yz}
\end{figure}

The difference by taking into account the vertical dispersion becomes apparent when examining the side-views seen in \cref{fig:lclustered-yz,fig:lartificial-5-yz,fig:lartificial-17-yz}, where the original trajectory data and the two version of representative trajectories are displayed. It clearly shows that modelling the vertical dispersion, actually \emph{represents} the vertical dispersion as seen in the original data.


\begin{figure}
\centering
    \includegraphics[width=\singlepic]{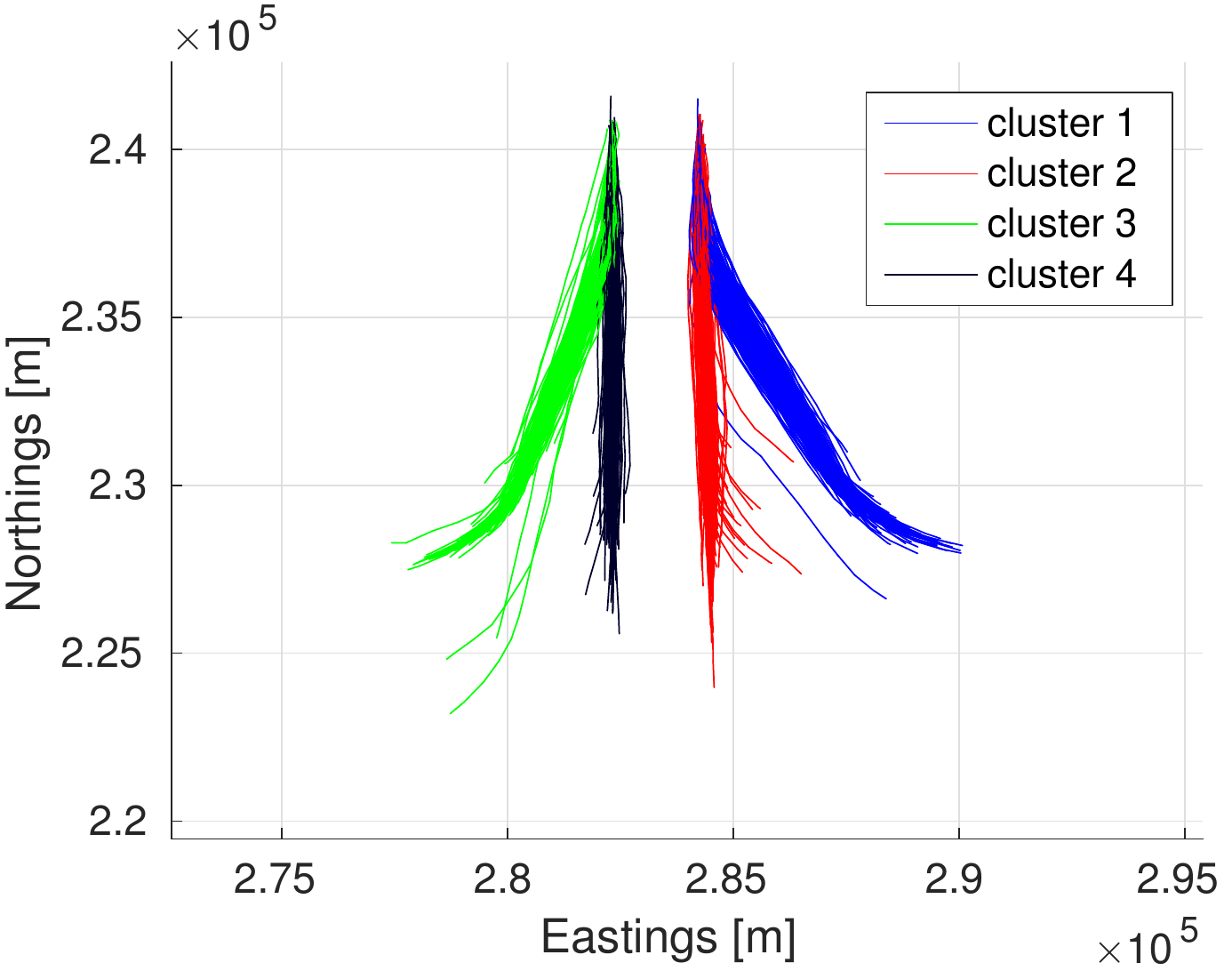}
    \caption{Top-view of all the clustered trajectories.}
    \label{fig:clustered-xy}
\end{figure}

For the evaluation a $100 \times 100$ grid is placed near the airport at a ground level. For each of these grid-points the percentage of the $815$ trajectories in the data that come within $300$ metres is calculated.
This simulates a simplistic approach to calculate from which locations the aircraft are vulnerable to an attack using \gls{rpgs}, without taking into account any mapping nor specific weapon characteristics.
For comparison, the analysis is performed on the original trajectory and the two versions of the representative trajectories.

\begin{figure}
\centering
    \includegraphics[width=\singlepic]{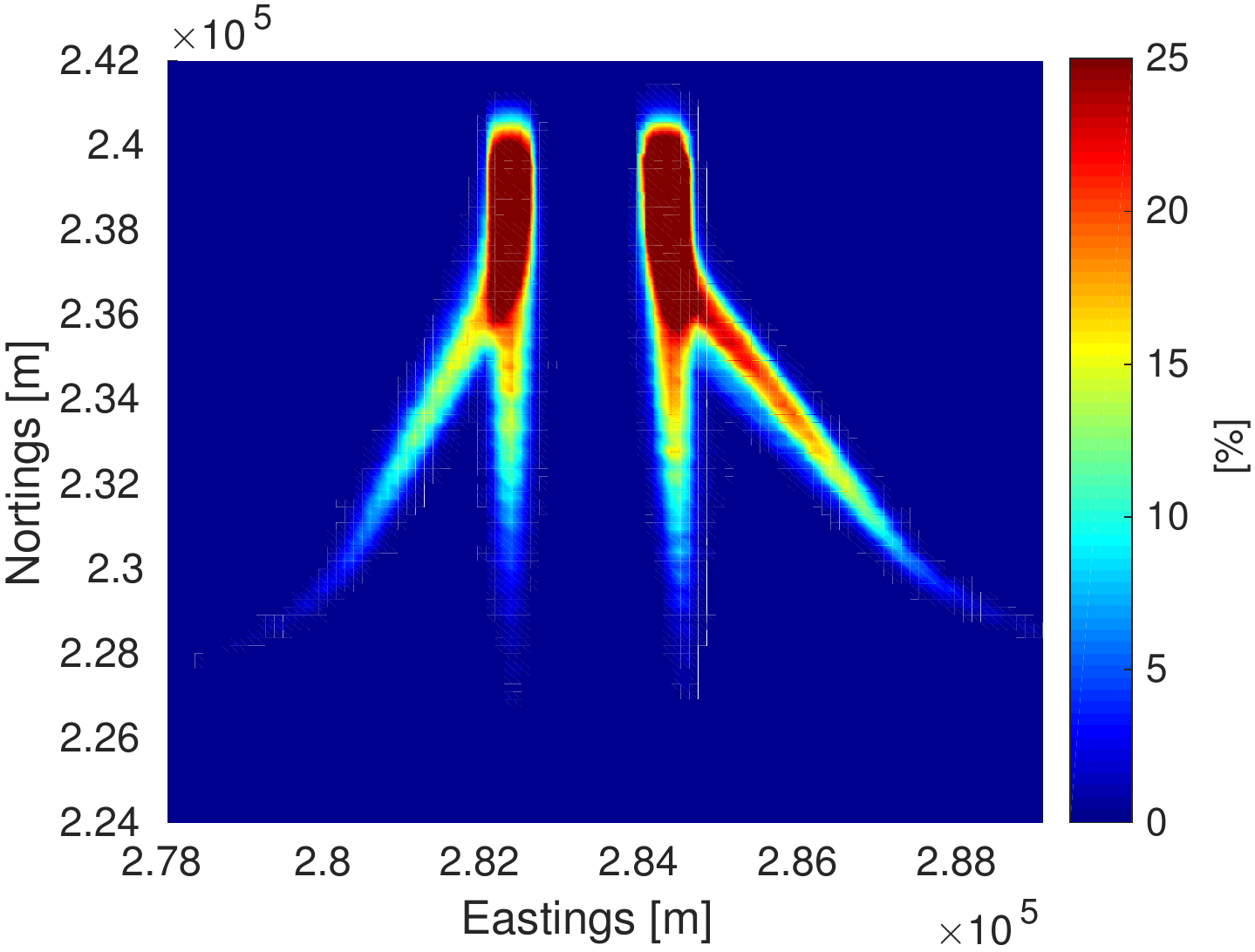}
    \caption{Footprint generated using the original trajectory data.}
    \label{fig:footprint-real}
\end{figure}

\begin{figure}
\centering
    \includegraphics[width=\singlepic]{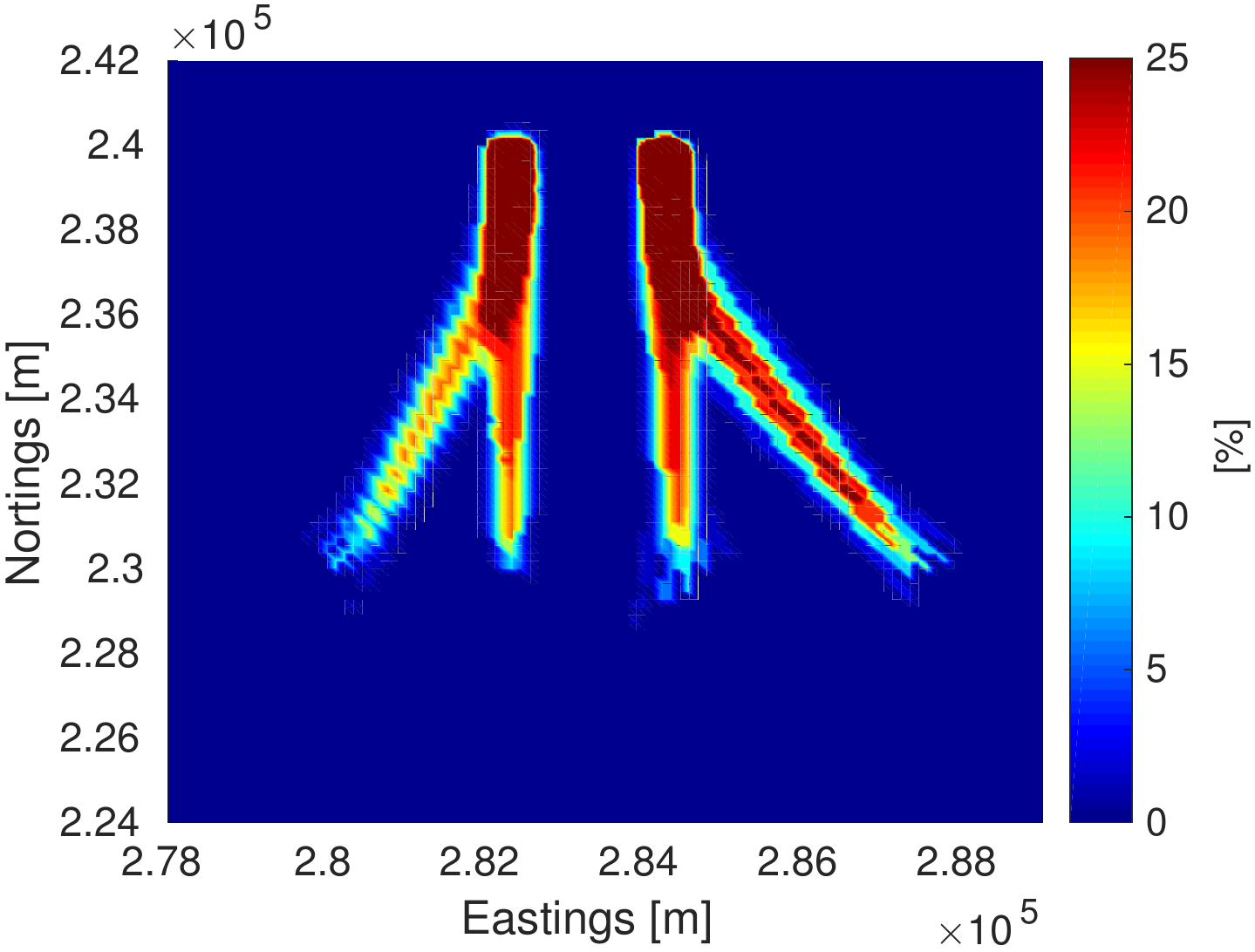}
    \caption{Footprint generated using representative trajectories (lateral dispersion only).}
    \label{fig:footprint-5}
\end{figure}

\begin{figure}
\centering
    \includegraphics[width=\singlepic]{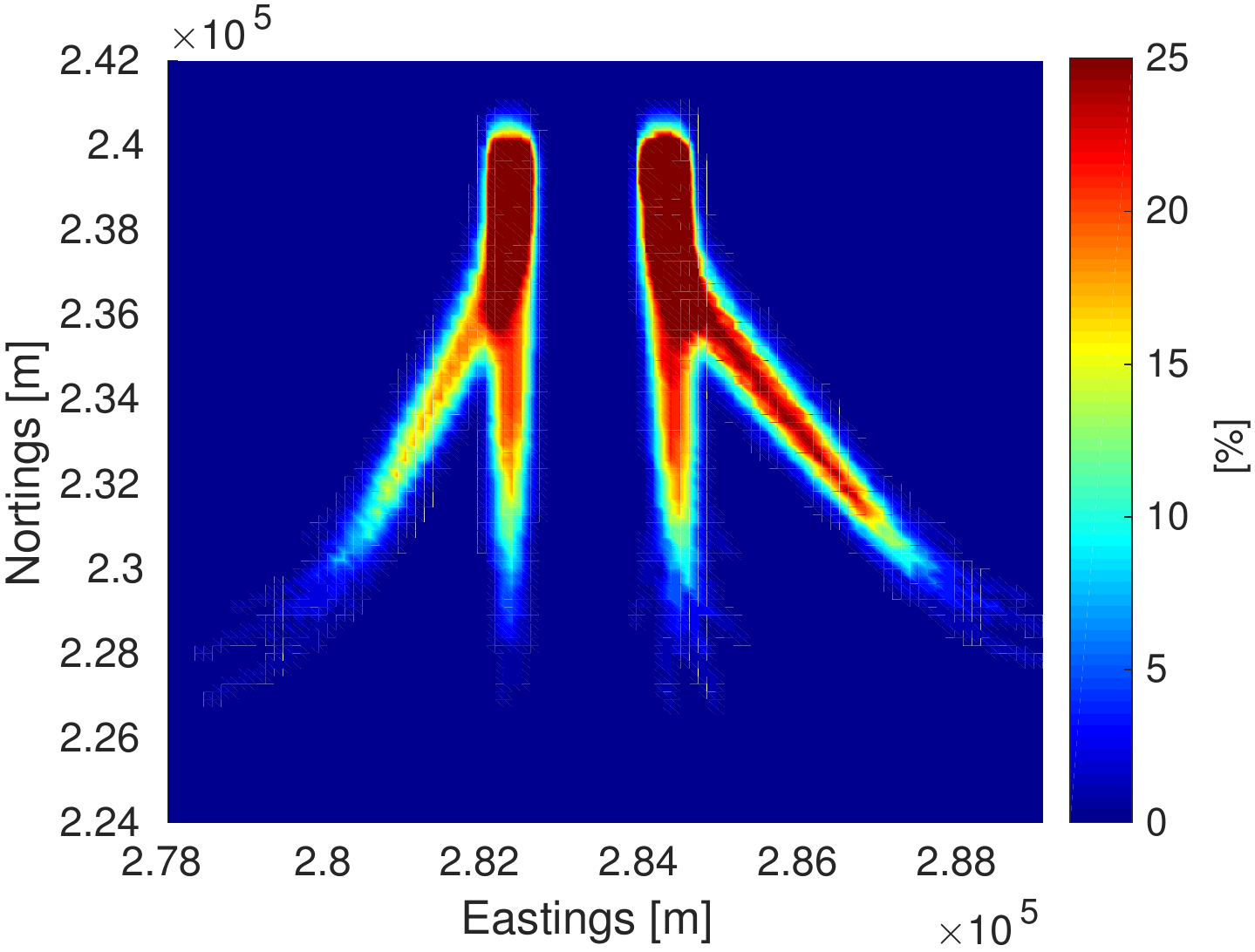}
    \caption{Footprint generated using representative trajectories (lateral and vertical dispersion).}
    \label{fig:footprint-17}
\end{figure}

The resulting footprint calculated for the original trajectory data is visible in \cref{fig:footprint-real}, where the range of percentages has been set to visualise the sweep of low percentages towards the end.
This sweep is missing from \cref{fig:footprint-5}, where only the lateral dispersion is taken into account. While the visible roughness is a direct effect of only using $5$ trajectories to represent each cluster, the missing sweep is a direct result of missing trajectories at a lower altitude.
As such, this sweep is visible in \cref{fig:footprint-17}. Here there are representative trajectories at a lower altitude, even at a distance further away from the airport.

The quantitative results comparing the methods to generate representative trajectories have been gathered in \cref{tab:footprint}. While the complete evaluated grid is $100 \times 100$, only the grid-points that have a non-zero value are analysed. For this reason the \emph{lateral dispersion only} has $2458$ grid-points and the \emph{lateral and vertical dispersion} has $2586$, as the latter extends further beyond the influence of the original trajectory data.
It demonstrates that including the vertical dispersion is beneficial for both spectrum of the extremes. Furthermore, the balance between underestimation and overestimation is included. Finally, the percentage of grid-points being under- and overestimated by more than $5\%$ is reported. This shows that less than $0.2\%$ of the \emph{active area} (i.e. non-zero grid-points) is underestimated by more than $5\%$. From a safety perspective, this is exactly the value we want as low as possible.
Interesting is also that $14.08\%$ is overestimated more than $5\%$ for the lateral and vertical dispersion, while $24.74\%$ is overestimated when only modelling the lateral dispersion. This can be interpreted as the vertical dispersion expanding the area of influence, yet reducing the intensity of the peaks.

\begin{table}
\centering
\caption{Comparing the footprints of the representative trajectories with the original trajectory data.} \label{tab:footprint}
\begin{tabular}{ | c | c | c | }
 \hline
    &  \textbf{lateral dispersion only} & \textbf{lateral and vertical dispersion} \\
  \hline
  minimum deviation                     & $-23.80 \%$   & $-8.30 \%$ \\
  maximum deviation                     & $24.53 \%$    & $17.52 \%$ \\
  grid-points underestimated            & $39.34 \%$ ($967$ out of $2458$)  & $28.15 \%$  ($728$ out of $2586$) \\
  grid-points overestimated             & $60.66 \%$ ($1491$ out of $2458$) & $71.85 \%$  ($1858$ out of $2586$) \\
  grid-points underestimated ($>5\%$)   & $1.42 \%$ ($35$ out of $2458$)    & $0.19 \%$   ($5$ out of $2586$) \\
  grid-points overestimated ($>5\%$)    & $24.74 \%$ ($608$ out of $2458$)  & $14.08 \%$  ($364$ out of $2586$) \\
  \hline
\end{tabular}
\end{table}


\section{Conclusions} \label{sec:conclusion}
The step-by-step method presented in this paper has shown the potential to capture the dispersion in a large historical data-set of trajectories using representative trajectories. Less than $0.2\%$ of the evaluations are underestimated by more than $5\%$, demonstrating its usefulness when performing a strategic analysis with the focus on safeguarding the airspace infrastructure. This while the number of trajectories has been diminished to $8.34\%$ of the original situation ($68$ versus $815$ trajectories). The strength of this technique is that the original data-set of trajectories can be increased, while the computational cost of the sequential calculations remain constant, and at the same time, retains its integrity.

Furthermore, while not explicitly demonstrated in this paper, it's possible to select key trajectories based on their location. E.g. when considering a \emph{hit} or \emph{miss} scenario concerning aircraft trajectories, a `bottom belly' of trajectories can be selected. The reasoning here would be, if the trajectories at a high altitude can be hit, it is a sure hit for trajectories at a lower altitude. This strength is directly connected to the ability to construct representative trajectories at any angle. The resulting representative trajectory can then matched (based on similarity) with a trajectory found in the historical data.

Finally, while the generation of a noise footprint has been mentioned on several occasions, and the results indicate that the extreme peaks are reduced and the lower levels expanded, no analysis has been performed concerning noise footprints. Such an analysis would have to include a noise model, as the noise as perceived from the ground will have a logarithmic effect. Nevertheless, it would be interesting to see how the inclusion of vertical dispersion influences the results concerning a noise footprint.

\section*{Acknowledgements} \label{sec:acknowledgements}
The authors gratefully acknowledge the funding provided under research grant EP/L505067/1 and industry sponsor Cunning Running Software Ltd.

\bibliography{wjeerland-scitech}
\bibliographystyle{abbrv}

\end{document}

%% file: glossary.tex
\newglossaryentry{m}
{
  name={\ensuremath{\mathbf{m}}},
  description={mean function},
  sort=m
}

\newglossaryentry{k}
{
  name={\ensuremath{\mathbf{k}}},
  description={covariance kernel},
  sort=k
}

\newglossaryentry{beta}
{
  name={\ensuremath{\beta}},
  description={precision parameter},
  sort=beta
}

\newglossaryentry{std}
{
  name={\ensuremath{\sigma}},
  description={standard deviation},
  sort={standard deviation}
}

\newglossaryentry{w}
{
  name={\ensuremath{\mathbf{w}}},
  description={single dimensions in parameter-space $\mathbb{R}^w$},
  sort=w
}

\newglossaryentry{wp}
{
  name={\ensuremath{\mathbf{W}}},
  description={multiple dimensions in parameter-space $\mathbb{R}^w$},
  sort=wp
}

\newglossaryentry{h}
{
  name={\ensuremath{\phi}},
  description={basis function for a scalar},
  sort=ht
}

\newglossaryentry{hv}
{
  name={\ensuremath{\boldsymbol{\phi}}},
  description={linear basis},
  sort=hv
}

\newglossaryentry{H}
{
  name={\ensuremath{\boldsymbol{\phi}}},
  description={basis functions for a vector},
  sort=Ht
}

\newglossaryentry{Hp}
{
  name={\ensuremath{\boldsymbol{\Phi}}},
  description={a block-diagonal matrix consisting of basis functions},
  sort=Hstar
}

\newglossaryentry{Y}
{
  name={\ensuremath{\mathbf{Y}}},
  description={a matrix describing a multi-dimensional trajectory},
  sort=T
}

\newglossaryentry{mw}
{
  name={\ensuremath{\boldsymbol{\mu}}},
  description={mean vector of parameter-space},
  sort=mw
}

\newglossaryentry{mf}
{
  name={\ensuremath{\boldsymbol{\mu}^\mathbf{f}}},
  description={mean vector of real-space},
  sort=mf
}

\newglossaryentry{sw}
{
  name={\ensuremath{\boldsymbol{\Sigma}}},
  description={co-variance matrix of parameter-space},
  sort=sw
}

\newglossaryentry{sf}
{
  name={\ensuremath{\boldsymbol{\Sigma}^\mathbf{f}}},
  description={co-variance matrix of real-space},
  sort=sf
}

\newglossaryentry{tv}
{
  name={\ensuremath{\boldsymbol{\tau}}},
  description={normalized time vector},
  sort=tauv
}

\newglossaryentry{yv} 
{
  name={\ensuremath{\mathbf{y}}},
  description={single trajectory},
  sort=y
}

\newglossaryentry{y} 
{
  name={\ensuremath{y}},
  description={single trajectory},
  sort=y
}

\newglossaryentry{xv} 
{
  name={\ensuremath{\mathbf{x}}},
  description={single dimension},
  sort=x
}

\newglossaryentry{x}
{
  name={\ensuremath{x}},
  description={single point},
  sort=x
}

\newglossaryentry{timev}
{
  name={\ensuremath{\mathbf{t}}},
  description={time vector},
  sort=tauv
}

\newglossaryentry{time}
{
  name={\ensuremath{t}},
  description={time},
  sort=tau
}

\newglossaryentry{t}
{
  name={\ensuremath{\tau}},
  description={normalized time},
  sort=tau
}

\newglossaryentry{f}
{
  name={\ensuremath{\mathbf{f}}},
  description={real-space},
  sort=f
}

\newglossaryentry{eps}
{
  name={\ensuremath{\epsilon}},
  description={$\epsilon$-neighborhood parameter of the DBSCAN algorithm},
  sort=epsilon
}

\newglossaryentry{N}
{
  name={\ensuremath{\mathcal{N}}},
  description={Gaussian function},
  sort=N
}

\newglossaryentry{O}
{
  name={\ensuremath{\mathcal{O}}},
  description={order of magnitude},
  sort=O
}

\newglossaryentry{E}
{
  name={\ensuremath{\mathbb{E}}},
  description={Expectation},
  sort=E
}

\newglossaryentry{I}
{
  name={\ensuremath{\mathbf{I}}},
  description={identity matrix},
  sort=I
}

\newglossaryentry{L}
{
  name={\ensuremath{\mathcal{L}}},
  description={Likelihood function},
  sort=L
}

\newglossaryentry{mu}
{
  name={\ensuremath{\boldsymbol{\mu}}},
  description={mean vector},
  sort=mu
}

\newglossaryentry{sigma1}
{
  name={\ensuremath{\mathbf{S}}},
  description={covariance matrix of an individual trajectory},
  sort=S
}

\newglossaryentry{sigma}
{
  name={\ensuremath{\boldsymbol{\Sigma}}},
  description={covariance matrix},
  sort=sigma
}

\newglossaryentry{var}
{
  name={\ensuremath{\sigma^2}},
  description={variance},
  sort={variance}
}

\newglossaryentry{muh}
{
  name={\ensuremath{\boldsymbol{\hat \mu}}},
  description={mean vector estimate},
  sort=muh
}

\newglossaryentry{sigmah}
{
  name={\ensuremath{\boldsymbol{\hat \Sigma}}},
  description={covariance matrix estimate},
  sort=sigmah
}

\newglossaryentry{betah}
{
  name={\ensuremath{\hat \beta}},
  description={precision parameter estimate},
  sort=betah
}

\newglossaryentry{wv}
{
  name={\ensuremath{\mathbf{w}}},
  description={weights vector},
  sort=wv
}

\newacronym{gp}{GP}{Gaussian Process}
\newacronym{gps}{GPs}{Gaussian Processes}
\newacronym{gpr}{GPR}{Gaussian Process Regression}
\newacronym{gmr}{GMR}{Gaussian Mixture Regression}
\newacronym{ks}{K-S}{Kolmogorov-Smirnov}
\newacronym{em}{EM}{Expectation-Maximisation}
\newacronym{pca}{PCA}{Principle Component Analysis}
\newacronym{ecdf}{ECDF}{Empirical Cumulative Distribution Function}
\newacronym{cdf}{CDF}{Cumulative Distribution Function}
\newacronym{inm}{INM}{Integrated Noise Model}
\newacronym{atc}{ATC}{Air Traffic Control}
\newacronym{dft}{DFT}{Discrete Fourier Transform}
\newacronym{dbs}{DBSCAN}{Density-Based Spatial Clustering of Applications}
\newacronym{dtw}{DTW}{Dynamic Time Warping}
\newacronym{mcmc}{MCMC}{Markov Chain Monte Carlo}
\newacronym{rpgs}{RPGs}{Rocket-Propelled Grenades}
\newacronym{rpg}{RPG}{Rocket-Propelled Grenade}